# Accelerated System-Reliability-based Disaster Resilience Analysis for Structural Systems

Taeyong Kim[1] and Sang-ri Yi[2*]

[1] Department of Civil Systems Engineering, Ajou University, Suwon, Republic of Korea
[2] Department of Civil and Environmental Engineering, University of California Berkeley, Berkeley, USA
[*] Correspondence: yisangri@berkeley.edu

**Abstract:** Resilience has emerged as a crucial concept for evaluating structural performance under disasters because of its ability to extend beyond traditional risk assessments, accounting for a system's ability to minimize disruptions and maintain functionality during recovery. To facilitate the holistic understanding of resilience performance in structural systems, a system-reliability-based disaster resilience analysis framework was developed. The framework describes resilience using three criteria: reliability ($\beta$), redundancy ($\pi$), and recoverability ($\gamma$), and the system's internal resilience is evaluated by inspecting the characteristics of reliability and redundancy for different possible progressive failure modes. However, the practical application of this framework has been limited to complex structures with numerous sub-components, as it becomes intractable to evaluate the performances for all possible initial disruption scenarios. To bridge the gap between the theory and practical use, especially for evaluating reliability and redundancy, this study centers on the idea that the computational burden can be substantially alleviated by focusing on initial disruption scenarios that are practically significant. To achieve this research goal, we propose three methods to efficiently eliminate insignificant scenarios: the sequential search method, the *n*-ball sampling method, and the surrogate model-based adaptive sampling algorithm. Three numerical examples, including buildings and a bridge, are introduced to prove the applicability and efficiency of the proposed approaches. The findings of this study are expected to offer practical solutions to the challenges of assessing resilience performance in complex structural systems.

## 1. Introduction

Resilience has gained increased attention in civil engineering, particularly in the context of disaster risk management. While the traditional risk assessment framework focuses solely on the immediate safety and economic losses of structural systems after disasters, resilience additionally recognizes that structural systems are vulnerable to a range of disasters and that failures can have far-reaching consequences. Therefore, the concept of resilience necessitates the consideration of additional factors, such as the ability to maintain essential functions after a member disruption and to recover quickly and effectively to the original equilibrium states.

With the growing interest in the concept of resilience, there has been a surge in research aimed at developing new tools and methods for assessing the resilience of systems. One of the most widely used frameworks is proposed by Bruneau et al. (2003). This framework conceptually defines resilience using "4R" - robustness, redundancy, resourcefulness, and rapidity - and quantifies the performance of a system by estimating the loss and recovery of its functionality over time. This method is commonly referred to as the resilience triangle framework, as the upper area of the recovery curve forms a triangular shape. By leveraging this framework, researchers have proposed methods for assessing the resilience performance in different scales of interest, e.g., individual structures (Hosseini et al., 2016; Jiang et al., 2020), lifeline networks (Goldbeck et al., 2019; Han et al., 2021), and urban communities (Mayer, 2019; Rus et al., 2018). Additionally, various numerical techniques have been introduced to quantify the resilience performance including the sampling method (Burton et al., 2019; Wang & van de Lindt, 2021), optimization-based approach (Hernandez et al., 2014), heuristic methodologies (Dessavre et al., 2016; González et al., 2016). Such research efforts have contributed significantly to our understanding of resilience and provided valuable insights to enhance the resilience performance of civil infrastructure systems.

Another resilience analysis framework has been developed based on a system-reliability perspective, proposed by Lim et al. (2022) for static loads and further developed by Yi & Kim (2023) for stochastic loads. This approach defines resilience using "3R" - reliability, redundancy, and recoverability - and describes the role of each criterion to three different scales of civil infrastructures – individual structures, lifeline networks, and urban communities. For the structural-level resilience assessment, the reliability and redundancy indices (denoted as $\beta$ and $\pi$) were proposed as basic indicators of the structure's *internal* resilience, which should be evaluated repeatedly for different possible initial disruption scenarios. For each disruption scenario, the two indices give information respectively on (i) the likelihood of the initial disruption occurrence, and (ii) the likelihood of the subsequent progressive failures. The introduction of a probabilistic resilience constraint from a *de minimis* risk viewpoint and a graphical tool called the $\beta$–$\pi$ diagram allows for the use of these indices to identify initial disruption scenarios that do not satisfy the disaster resilience threshold. This framework facilitates a better understanding of the system's performance, broadening the scope of research into resilience-based structural design and assessment.

However, the system-reliability-based resilience analysis framework inevitably involves a huge computational cost needing to calculate the two resilience indices ($\beta$ and $\pi$) for every initial disruption scenario. As described in Lim et al. (2022) and Yi & Kim (2023), the number of initial disruption scenarios grows exponentially with the number of structural components in the system, as will be illustrated in Sections 3 and 4. This issue is particularly critical for complex structural systems that include a large number of components, as the number of possible initial disruption scenarios to be considered in the framework may become intractably large.

In principle, there can be two primary ways to alleviate the computational burden in assessing resilience performance. The first approach involves employing advanced reliability analysis techniques to accelerate the calculation of each reliability and redundancy indices. Machine learning algorithms, for example, can be employed as a surrogate model to quickly estimate the target responses of complex structural systems (Song et al., 2021). Importance sampling techniques are another instance (Au & Beck, 2001; Ching & Chen, 2007; Hohenbichler & Rackwitz, 1988; Wang et al., 2019). The second approach, on the other hand, is to introduce a prescreening step to eliminate trivial scenarios from the analysis. This approach leverages the fact that the goal of the resilience assessment, whether it is for performance evaluation or the design of a structure, is to determine if each combination of reliability and redundancy satisfies the target resilience requirement. Therefore, if an algorithm can efficiently screen out trivial scenarios of which the reliability-redundancy indices satisfy the resilience requirement with a substantial margin, one can avoid performing a detailed time-consuming reliability-redundancy analysis for such scenarios. This study focuses on proposing methods for the latter approach and further shows in numerical examples that the efficiency can be amplified when the two approaches are employed simultaneously.

The study proposes three pre-screening methods, namely, the sequential search method, the *n*-ball sampling method, and the surrogate model-based adaptive sampling algorithm. The sequential search method identifies noteworthy, i.e., non-trivial, initial disruption scenarios by establishing an automated process to recursively update the upper bounds of the reliability index estimations. For this, the conceptual and mathematical definitions of noteworthy scenarios are newly proposed. The *n*-ball sampling method identifies noteworthy initial disruption scenarios by utilizing a random search technique in an *n*-dimensional hypersphere called *n*-ball. Lastly, the surrogate model-based adaptive sampling algorithm leverages a machine learning-based surrogate model to reduce the number of simulation runs required in the *n*-ball sampling method. To provide a clear understanding, the methods are explained with an illustrative example of a two-layer Daniels system (Daniels, 1945).

Meanwhile, acknowledging the heuristic nature of the latter two approaches (*n*-ball sampling and surrogate-based adaptive sampling), i.e., the methods are not rigorously guaranteed to always identify all noteworthy scenarios, the algorithms are conservatively designed. An extensive investigation and discussion on their robustness aspect is provided through a series of numerical examples.

The paper comprises five sections. In Section 2, a high-level overview of the system-reliability-based resilience analysis framework is provided, including the definitions of reliability and redundancy indices and the resilience threshold. Section 3 introduces three novel approaches designed to accelerate resilience analysis for structural systems. These methods include the sequential search method, the *n*-ball sampling method, and the surrogate model-based adaptive algorithm. Section 4 investigates four numerical examples to demonstrate the efficiency and applicability of the proposed methods. Finally, Section 5 presents a summary of the paper's main findings and conclusions.

**2. System-Reliability-based Disaster Resilience Analysis**

2.1. Reliability and redundancy indices

According to Lim et al. (2022), the resilience of a system is described in terms of different progressive failure modes, represented as a collection of *initial disruption scenarios*. Reliability, in this context, refers to the structural system's ability to resist the initial component disruptions caused by external loads, while redundancy denotes the system's capacity to prevent progressive failures even after some structural components have failed. Each capacity is expressed as the probability of occurrence under the uncertainties in disaster and systems characteristics. In other words, the reliability index $\beta_{i,j}$ for an initial disruption scenario $F_i$ and a hazard event $H_j$ is defined as

$$\beta_{i,j} = -\Phi^{-1}\left(P(F_i|H_j)\right) \tag{1}$$

where $\Phi^{-1}(\cdot)$ denotes the inverse cumulative distribution function (CDF) of the standard Gaussian distribution. Similarly, the corresponding redundancy index $\pi_{i,j}$ is given as

$$\pi_{i,j} = -\Phi^{-1}\left(P(F_{sys}|F_i, H_j)\right) \tag{2}$$

in which $F_{sys}$ denotes the system-level failure of the given structure. Given a condition that the initial disruption scenarios are mutually exclusive and collectively exhaustive (MECE) events, the annual failure probability of structural systems under the hazard $H_j$, $P(F_{sys,j})$ can be written using the reliability and redundancy indices as

$$P(F_{sys,j}) = \sum_i P(F_{sys,i,j}) = \sum_i P(F_{sys}|F_i, H_j) P(F_i|H_j) \lambda_{H_j} = \sum_i \Phi(-\pi_{i,j}) \Phi(-\beta_{i,j}) \lambda_{H_j} \tag{3}$$

Here, $P(F_{sys,i,j})$ represents the annual system failure probability originating from $F_i$ under $H_j$, and $\lambda_{H_j}$ is the annual mean occurrence rate of $H_j$. The MECE initial disruption scenarios can be defined in terms of the component failure events using the set theory (Yi & Kim (2023)), which can be found in Section 3.2. The reliability and redundancy indices respectively represent the capability of components and system against the given disruption scenario and, at the same time, can jointly describe the system's failure probability via Eq. (3).

2.2. Resilience threshold

To systemically evaluate the resilience performance of structural systems using the indices proposed in Eqs. (1) and (2), Lim et al. (2022) introduced the concept of *de minimis* risk. In engineering risk management, *de minimis* refers to a risk level that is considered trivial or minor and do not merit further consideration. The overall resilience of the system can be secured by introducing an upper bound of $P(F_{sys,i,j})$ as (Yi & Kim, 2023)

$$P(F_{sys,i,j}) = \Phi(-\pi_{i,j}) \Phi(-\beta_{i,j}) \lambda_{H_j} < P_{dm}/N_F \tag{4}$$

where $P_{dm}$ represents the *de minimis* risk which is typically on the order of $10^{-7}/yr$ for the civil structural systems (Paté-Cornell, 1994), and $N_F$ stands for the number of initial disruption scenarios considered in the analysis. Dividing Eq. (4) by $\lambda_{H_j}$, the equation can be written as

$$\Phi(-\pi_{i,j})\Phi(-\beta_{i,j}) < P_{dm}/(\lambda_{H_j}N_F) \qquad (5)$$

where $P_{dm}/(\lambda_{H_j}N_F)$ is used as a criterion to determine resilience status of each initial disruption scenario.

To utilize the assessed resilience status in decision-making, Lim et al. (2022) proposed a graphical tool called a β–π diagram, illustrating both resilience indices and the resilience threshold in Eq. (5) as shown in Figure 1. In the figure, the blue circles represent the pairs of resilience indices for each initial disruption scenario, while an orange dashed line represents the resilience threshold. The circles located inside the orange area correspond to scenarios that lack the required level of resilience. This diagram not only facilitates the identification of potentially critical disruption scenarios that do not meet the target resilience performance but also provides insights into the disaster resilience of structural systems by offering a compact representation of different system failure paths.

The diagram is also useful in the decision-making process (Lim et al., 2023). For instance, in achieving the required level of disaster resilience, preventive actions can be carried out to push the marker beyond the resilience threshold (depicted as a dashed line in Figure 1) by improving the reliability and/or redundancy of relevant scenarios. Enhancing reliability typically involves substituting a structural component with a 'stronger' alternative, while increasing redundancy can be achieved through various means, such as integrating additional structural elements. Given various ways available to meet the targeted resilience level, it is necessary to develop a new framework incorporating appropriate constraints and optimization algorithms featuring the core of the resilience-based design optimization.

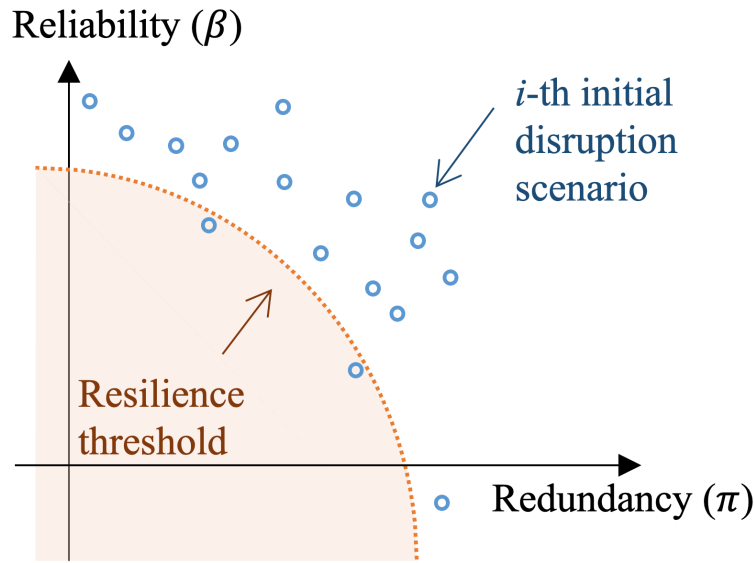

**Figure 1.** Illustration of β–π diagram

**3. Approaches to Accelerate the System-Reliability-based Resilience Analysis**

Three approaches are proposed to address the computational challenges in the system-reliability-based resilience analysis. The challenges, herein, represent the computational burden in estimating the reliability and redundancy indices for every initial disruption scenario. To effectively explain the proposed methods, this section begins by providing the details of the illustrative example and the final β–π diagram obtained using the conventional approach, relying on naïve Monte Carlo simulation. Then, the three new methods are presented to accelerate the procedure to obtain the β–π diagram.

3.1. Illustrative example

To illustrate the proposed concepts, we introduce a Daniels system (Daniels, 1945), which has been widely used in various system reliability problems (Gollwitzer & Rackwitz, 1990; Lim et al., 2022; Song & Der Kiureghian, 2003). The system consists of five perfectly brittle bars arranged in a double-layer structure, as shown in Figure 2. Bars 1 and 2 have a cross-sectional area of 1.5 mm², while bars 3, 4, and 5 have cross-sectional areas of 2 mm², 1 mm², and 1 mm², respectively. The yield stress of the bars is assumed to follow a Gaussian distribution with a mean of 400 MPa, and a coefficient of variation (c.o.v) of 0.3, 0.1, 0.35, 0.2, and 0.15 for bars 1, 2, 3, 4, and 5, respectively. The Daniels system is subjected to a vertical load representing a hazard of interest $H_j$ in Eq. (1), with a scalar value of 600 N. For simplicity, we assume that all random variables are statistically independent of each other. More complex numerical examples will be presented in Section 4.

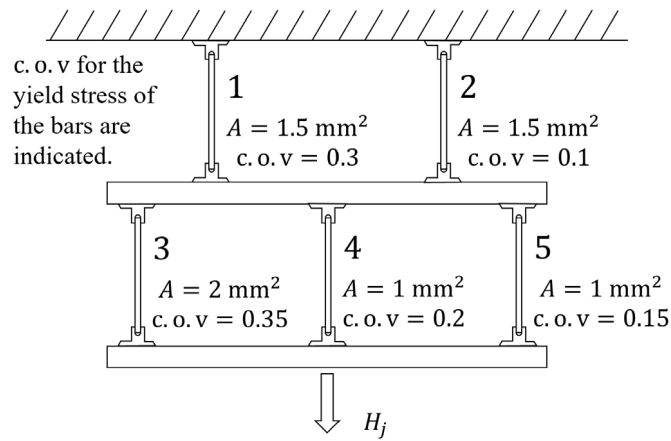

**Figure 2.** Double-layer Daniels system

The initial disruption scenarios ($F_i$) can be defined as all possible combinations of component failures that can occur within the system. Since the Daniels system consists of five structural components, one can define a total of 32 (= $2^5$) initial disruption scenarios. For each initial disruption scenario, reliability and redundancy indices are calculated using Eqs. (1) and (2), respectively and presented in Figure 3 along with a resilience threshold of $10^{-4}$. In the resilience analysis of the Daniels system, component failure is defined as the applied stress exceeding the yield stress, whereas system failure ($F_{sys}$) occurs when all bars in a layer fail. Various reliability analysis techniques can be applied (Song et al., 2022), but in this example, the reliability index is analytically derived in a closed-form expression, while Monte Carlo simulation (MCS) is employed to estimate the redundancy index.

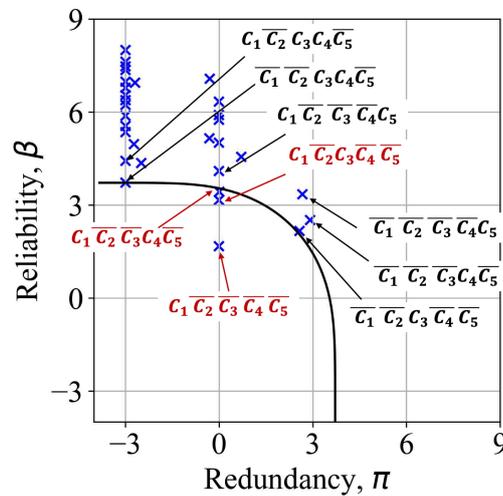

**Figure 3.** β–π diagram of the double-layer Daniels system

The number of initial disruption scenarios increases exponentially as the number of structural components in a system increases. Therefore, even when advanced reliability analysis techniques, such as adaptive algorithms, are introduced to expedite the computation of reliability and redundancy indices, the sheer volume of computations required for different scenarios can be highly demanding. One way of reducing computational cost is to first identify the initial disruption scenarios that fall within or adjacent to the resilience threshold and then focus the computational resources to precisely analyze only those scenarios.

For example, in Figure 3, the initial disruption scenarios that do not meet the target resilience level, therefore requiring detailed attention, are denoted in red next to the corresponding β–π point, where $C_i$ and $\overline{C_i}$ represent the failure and non-failure of $i$-th bar, respectively. The intersection notation ∩ is omitted in the labels. Additionally, the initial disruption scenarios that satisfy the resilience target, but are located near the resilience threshold are denoted in black. This study terms such scenarios (both denoted in red and black in Figure 3) as "noteworthy initial disruption scenarios." Among the list of noteworthy initial disruption scenarios, the scenarios that fail to meet the resilience threshold (e.g., scenarios denoted in red in Figure 3) can be identified based on Eq. (5).

In the subsequent subsections, three distinct approaches are introduced to rapidly identify the noteworthy scenarios. Note that this section focuses only on identifying the list of noteworthy scenarios, and details of the subsequent process of the workflow (i.e., using advanced reliability analysis techniques to identify reliability and redundancy indices of the noteworthy scenarios) are not covered. The overall efficiency will be demonstrated through numerical examples.

## 3.2. Sequential search method

Extending the subtraction method in Yi & Kim (2023), we propose a sequential search method to find noteworthy scenarios. The subtraction method is a systematic approach to relate the probability of initial disruption scenarios (denoted using the letter $F$) and the single component failure events (denoted using the letter $C$). To start with, let us denote $F$ in terms of the component failure events $C$ using the set theory (Yi & Kim, 2023):

$$\boldsymbol{F} = \{F | F = (\cap_{i \in \mathbf{S}} C_i) \cap (\cap_{j \in \mathbf{S}^c} \overline{C_j}), \mathbf{S} \subset \{1, 2, \ldots, N_c\}\} \tag{6}$$

where $\overline{C_j}$ denotes the survival of member $j$ and $\mathbf{S}^c$ is the complement set of $\mathbf{S}$. Note that this definition of $\boldsymbol{F}$ always satisfies MECE condition.

The "noteworthiness" is evaluated based on two main principles. First, if the following condition is satisfied for the $\beta_i$,

$$\Phi(-\beta_i) < P_{dm}/(\lambda_H N_F) \tag{7}$$

such a scenario $F_i$ is classified as a trivial scenario, and no further analysis is required to calculate $\pi_i$, as the resilience performance in Eq. (5) is automatically satisfied regardless of the values of $\pi_i$. Second, if an initial disruption scenario $F_i$ is a subcase of a component failure scenario $C_k$ ($F_i \subset C_k$) and the value of $\beta$ for the component failure $C_k$ (i.e., $F_i$ in Eq. (1) is replaced by component failure event $C_k$) satisfies Eq. (7), the scenario is also considered a trivial scenario. This is because the probability of $F_i$ is always smaller than $C_k$. This principle applies not only to single component failure scenarios $C_k$ but also to joint components failure scenarios such as $C_k C_l (k \neq l)$, $C_k C_l C_m (k \neq l, k \neq m, l \neq m)$, and so on, where $k, l$ and $m$ stand for the index of structural components within the system.

Based on these two screening principles, we propose a systematic search approach to identify noteworthy scenarios. The search process commences with the calculation of the reliability index for the failure of each member shown in Phase 1 of Figure 4, where the calculated reliability index is also indicated. The reliability index is then compared with the resilience threshold to identify components that can be excluded in the subsequent phases. For instance, in this example, we found that only the reliability of $C_2$ is lower than the threshold risk level $10^{-4}$ (Note, $-\Phi^{-1}(10^{-4}) = 3.72$). The selected node is highlighted using a dashed circle in Figure 4.

In the next phase (Phase 2), the same process is repeated for events involving the failure of two members. However, the components already eliminated (indicated by the dashed circle in Phase 1) in the previous phase are not considered in the current and future phases. The reliability index of each joint failure event is computed and compared with the resilience threshold. In the double-layer Daniels system, six events are evaluated in Phase 2, excluding the joint events that involve the failure of bar 2. Four events ($C_1C_5$, $C_3C_4$, $C_3C_5$, $C_4C_5$) are eventually selected as excluded events as highlighted using dashed circles in Phase 2 of Figure 4.

The process of identifying excluded events is repeated until no further decomposition is necessary. Specifically, in this example, the process is terminated in Phase 2 because there are no intersecting events that do not include any of the identified "excluded events" ($C_2$, $C_1C_5$, $C_3C_4$, $C_3C_5$, $C_4C_5$). Using the list of excluded events identified from the search technique, the final noteworthy initial failure scenarios can be determined. Among all possible initial disruption scenarios, those that are not a subset of the identified excluded events are selected and listed in Table 1. Then, the reliability and redundancy indices are calculated for the filtered events to assess whether the combination of reliability and redundancy performance satisfies the target resilience requirement.

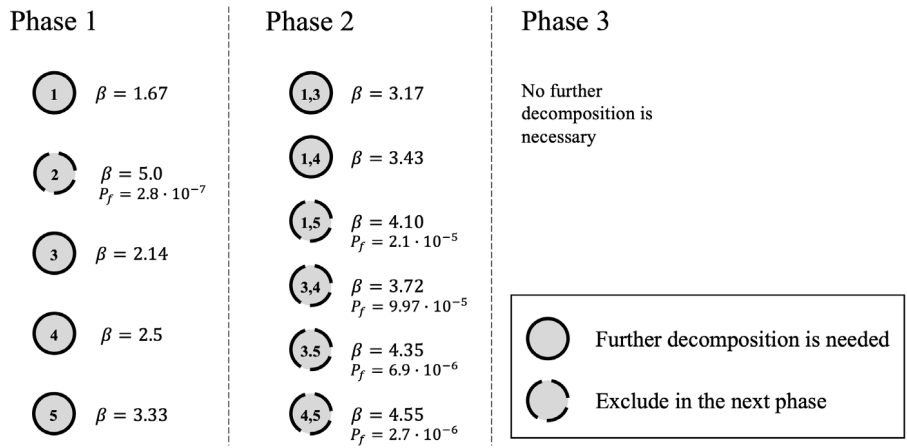

**Figure 4.** The sequential search method applied to the double-layer Daniels system

In the example, the proposed method successfully identified six noteworthy initial disruption scenarios out of 32 MECE events. The reliability and redundancy of the possible noteworthy scenarios are then evaluated. Table 1 shows that the first three cases (Cases 1, 2, and 3) are identified as scenarios that fail to satisfy the resilience threshold, which aligns with the results presented in Figure 3. Since the proposed search method conservatively selects the noteworthy scenarios, three additional scenarios are also identified in Table 1 even though their final resilience performance satisfies the target resilience (Cases 4, 5, and 6). Note that the number of additional noteworthy scenarios that already satisfy the resilience threshold depends on the number of points located near the resilience threshold (black solid line in Figure 3). Meanwhile, understanding that conservativeness in the selection of noteworthy scenarios can hinder the effectiveness of prescreening and overall efficiency, a further parametric study is desired to understand the relationship between the effectiveness of prescreening and the configuration of structural systems and hazards loads. Nevertheless, in all inspected case studies, the proposed methods significantly reduced the total number of model evaluations in evaluating the reliability and redundancy indices even though it conservatively selected the noteworthy scenarios.

**Table 1.** Noteworthy scenarios identified by sequential search method

| Case | Disruption scenarios ($F_i$) | $\beta_i$ | $\pi_i$ | $-\Phi^{-1}(\Phi(-\beta_i)\Phi(-\pi_i))$ |
|---|---|---|---|---|
| 1 | $\{C_1\overline{C_2}C_3\overline{C_4}C_5\}$ | 3.17 | −0.01 | 3.36 |
| 2 | $\{C_1\overline{C_2}\overline{C_3}C_4C_5\}$ | 3.44 | 0.002 | 3.62 |
| 3 | $\{C_1\overline{C_2}C_3\overline{C_4}\overline{C_5}\}$ | 1.68 | −0.002 | 1.99 |
| 4 | $\{\overline{C_1}C_2C_3\overline{C_4}C_5\}$ | 2.17 | 2.57 | 3.79 |
| 5 | $\{\overline{C_1}C_2C_3C_4\overline{C_5}\}$ | 2.52 | 2.91 | 4.26 |
| 6 | $\{\overline{C_1}C_2C_3C_4C_5\}$ | 3.35 | 2.67 | 4.67 |

3.3. *n*-ball sampling method

Structural reliability problems are often challenging to solve using analytical methods, especially when the limit-state function is complex and difficult to differentiate (T. Kim & Song, 2018). In such cases, sampling-based methods becomes an attractive alternative. The *n*-ball sampling approach is a simulation-based method designed to identify noteworthy scenarios, where the noteworthiness is defined using Eq. (7).

To achieve this, the *n*-ball sampling method generates the samples of random variables uniformly over the interior of an *n*-dimensional hypersphere of radius $R$, called *n*-ball, where the sampling is processed in the standard normal space **u**. The corresponding values in the original space **x** are obtained by the probabilistic transform, $\mathbf{x} = \mathbf{T}^{-1}(\mathbf{u})$ such as inverse Nataf transformation (Der Kiureghian, 2022). To reduce the sampling variability, a Latin hypercube Sampling (LHS) technique (McKay et al., 2000) can be employed.

The radius of the *n*-ball should be determined to be able to capture all the noteworthy scenarios that violates Eq. (7). For example, setting radius of $R = -\Phi^{-1}(P_{dm}/\lambda_H N_F)$ covers the exact domain of significance when the limit state function of the reliability analysis is linear, i.e., in such cases, the scenarios that do not satisfy Eq. (7) always have a design point located within the *n*-ball of radius $R$. However, to account for the cases where the limit-state surface is nonlinear and not smooth, particularly when it is concave to the origin at the design point, it is recommended to define the radius slightly larger than $R$. However, it is remarked that $F_i$ is in nature defined as a parallel system, implying that the failure domain of $F_i$ in the random variable space is the intersection of that of multiple $C_i$ or $\bar{C}_i$ events. This characteristic makes the limit state surface more likely to be convex to the origin. Owing to this property, a small increment in $R$ is typically sufficient to identify most of the noteworthy initial disruption scenarios.

Once the samples of **u** have been generated, deterministic structural analyses are performed to identify the samples exhibiting component failures. The failure scenarios that appeared at least once in the samples are categorized as noteworthy scenarios. The resilience performance of these scenarios is then evaluated by verifying whether each failure scenario satisfies Eq. (5).

For the same Daniels system example, we randomly generated 10,000 samples from the 5-dimensional hypersphere with a radius of $R = -\Phi^{-1}(10^{-4}) = 3.719$, and identified 7 possible noteworthy scenarios as summarized in Table 2. Note that, in J. Kim & Song (2020), the promoted initial population size was 50,000 for $R = 5$. In our research, we used a smaller population size, as our radius is smaller. Compared to the sequential search method presented in Table 1, this method conservatively identified one more scenario to be noteworthy, i.e., Case 7 in Table 2. However, once the reliability index is calculated for Case 7, one can notice that the redundancy index (the shaded cells in Table 2) for this case does not need to be calculated since the reliability index already satisfies the resilience threshold as in Eq. (7). Finally, by checking Eq. (5) for all possible scenarios, it is shown that three scenarios fail to meet the resilience requirement, confirming the consistency with the previous results.

**Table 2.** Noteworthy failure scenarios estimated by the *n*-ball sampling method

| Case | Disruption scenarios ($F_i$) | $\beta_i$ | $\pi_i$ | $-\Phi^{-1}\big(\Phi(-\beta_i)\Phi(-\pi_i)\big)$ |
|---|---|---|---|---|
| 1 | $\{C_1\bar{C}_2C_3\bar{C}_4\bar{C}_5\}$ | 3.17 | $-0.01$ | 3.36 |
| 2 | $\{C_1\bar{C}_2\bar{C}_3C_4\bar{C}_5\}$ | 3.44 | 0.002 | 3.62 |
| 3 | $\{C_1\bar{C}_2\bar{C}_3\bar{C}_4C_5\}$ | 1.68 | $-0.002$ | 1.99 |
| 4 | $\{\bar{C}_1\bar{C}_2\bar{C}_3\bar{C}_4\bar{C}_5\}$ | 2.17 | 2.57 | 3.79 |
| 5 | $\{\bar{C}_1C_2\bar{C}_3C_4\bar{C}_5\}$ | 2.52 | 2.91 | 4.26 |
| 6 | $\{\bar{C}_1C_2C_3\bar{C}_4\bar{C}_5\}$ | 3.35 | 2.67 | 4.67 |
| 7 | $\{\bar{C}_1C_2C_3C_4\bar{C}_5\}$ | 3.73 | $-3.37$ | 3.73 |

3.4. Surrogate model-based adaptive sampling algorithm

Although the *n*-ball sampling method is a convenient approach for identifying noteworthy initial disruption scenarios, a limitation is observed when the method is applied to a problem that entails a significant computational demand per analysis. To address this challenge, we propose an adaptive algorithm that requires a smaller set of simulation points to construct a surrogate model to be used in the *n*-ball sampling approach. The method is inspired by previous research efforts (D.-S. Kim et al., 2013; J. Kim & Song, 2020; T. Kim et al., 2020; McKay et al., 2000). In this study, we employ a deep neural network (DNN) model as a surrogate, as deep learning has shown unprecedented performance in terms of identifying the intricate pattern between input and output variables in diverse fields including business, medical, and engineering (LeCun et al., 2015). Of course, other machine learning algorithms, such as Gaussian Processes or Support Vector Machines, could also be employed.

The algorithm is summarized below, and implementation details are provided afterward. The flowchart is presented in Figure 5.

- **Step 1**: Generate a small set of initial samples over an *n*-ball of radius $R^*$ in the standard normal space **u**. To properly identify the failure domain of initial disruption scenarios located adjacent to the resilience threshold, $R^*$ is set to $1.1R = -1.1 \cdot \Phi^{-1}(P_{dm}/\lambda_H N_F)$. While any points can be considered as candidates for the initial set, this study uses uniformly distributed sample points on the surface of the *n*-ball as the initial training set. This bounds the domain of interest and minimizes the extrapolation.
- **Step 2**: Conduct structural analyses for the samples generated in the previous step to identify the components' failure. For each sample point, a binary vector is obtained, indicating whether the corresponding components have failed (0) or remain safe (1). The length of the binary vector corresponds to the number of structural components in the problem.
- **Step 3**: Train a DNN model using the sample points generated in Step 1 and their corresponding binary failure outcomes from Step 2. The sigmoid function is adopted as the activation function of the last layer of the DNN model to predict the binary states of the structural components. The cross-entropy loss function is employed to minimize the difference between the predicted output and the actual binary values. The trained DNN model can be used to predict the binary outcomes for new sample points in the **u**-space, eliminating the need for running expensive structural analyses.
- **Step 4**: Generate a large number of sample points randomly in the region between two radii, referred to as a sampling ring. The radii are increased gradually from the origin up to $R^*$ in order to systemically cover the entire area over the interior of the *n*-dimensional hypersphere. The component states (fail/safe) of the generated sample points are predicted using the trained DNN model in Step 3. Once the sampling ring reaches the radius of the hypersphere, the sampling is carried out over the entire domain of the *n*-ball.
- **Step 5**: Identify a sample point where the DNN model's prediction is weak (i.e., the predicted value is close to 0.5). This sample point is added to the existing training dataset, and the DNN model is retrained using the updated dataset. For this, structural analysis is performed for the newly added dataset to obtain the output value of the training point. The retraining process aims to improve the accuracy of the DNN model in predicting the component states. Moreover, the mutation and crossover techniques which are often employed in genetic algorithms can be implemented to further explore the failure domain.
- **Step 6**: Verify the convergence of the algorithm. To ensure the robustness of the proposed algorithm, two stopping criteria are introduced. First, the radii used in Step 4 must be sufficiently large to represent the resilience limit. Second, the convergence ratio, which is defined as the ratio of the number of failed sample points to the total number of sample points predicted by the DNN model, must be stable. To evaluate the convergence ratio, sample points are randomly generated in the **u**-space, and the ratios are compared between consecutive iterations. If any of the two criteria are not met, the algorithm proceeds to Step 3 and repeats the process until convergence is achieved.

- **Step 7**: Finally, once the convergence criteria are satisfied, the noteworthy scenarios can be identified by investigating the points used in training the DNN model (i.e., selected points in Step 5). Alternatively, noteworthy scenarios can also be predicted using the trained DNN model.

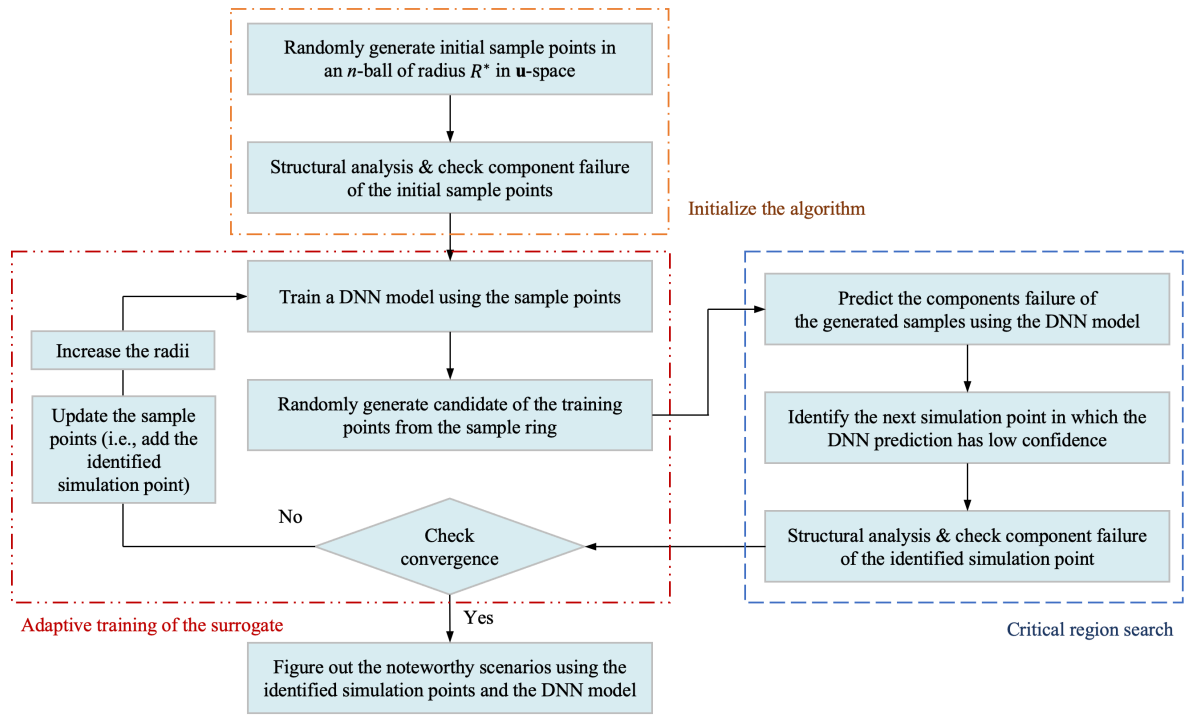

**Figure 5.** Flowchart of the surrogate model-based adaptive sampling method

A numerical investigation is carried out using the two-layer Daniels system to demonstrate the effectiveness of the surrogate model-based adaptive sampling method and provide a better understanding of the approach. The initial sample points are generated in the **u**-space by setting one of the components of the sample equal to the radius representing the resilience threshold, $R = -\Phi^{-1}(P_{dm}/\lambda_H N_F)$, while keeping all other components at zero. This practice makes the sample points be placed on the surface of hypersphere of radius $R$. Since the Daniels system takes 5 random variables, 10 initial sample points are generated, e.g., $[R, 0, 0, 0, 0]$, $[-R, 0, 0, 0, 0]$, and $[0, 0, 0, 0, -R]$. Then, the radius of the $n$-ball hypersphere, $R^*$ is discretized into $N_R$ steps, and the inside diameter of the sampling ring for the first iteration is set to $R^*/N_R$, which increases by $R^*/N_R$ for each iteration. The distance between the outside and inside radii of the sampling ring is set to $3R^*/N_R$. The gradual increment of radius is to uniformly populate the samples across different radius and $N_R = 35$ is selected in this example. However, it is necessary to conduct a parametric study to enhance the algorithm's performance further.

The architecture of the DNN surrogate model used in the algorithm is illustrated in Figure 6. Batch normalization is applied before the hidden layer. Moreover, distinct colors are used based on the activation function; a hidden layer with the Rectified Linear Unit (ReLU) is represented by green, whereas one with the sigmoid function is denoted by orange. The model is designed to predict the components' states of each sample point.

The number of layers and units are determined to avoid convergence to local optima. The training process starts with 300 epochs for the first iteration, and the number of epochs increased by 10 at each iteration to apply the simulation annealing effect. Note that an epoch refers to one complete iteration of feeding the entire training dataset through the DNN model. The maximum number of epochs is set to 500 to reduce the computational costs of training the DNN model. The Adam optimizer (Reddi et al., 2008) is used as the optimization algorithm to minimize the cross-entropy loss of the output with a batch size of 20, where the optimizer updates the weights and biases of the model based on the loss function's gradients. The iteration continues until the convergence criteria are met. The stopping criterion for the convergence ratio of consecutive iterations is set to 0.001.

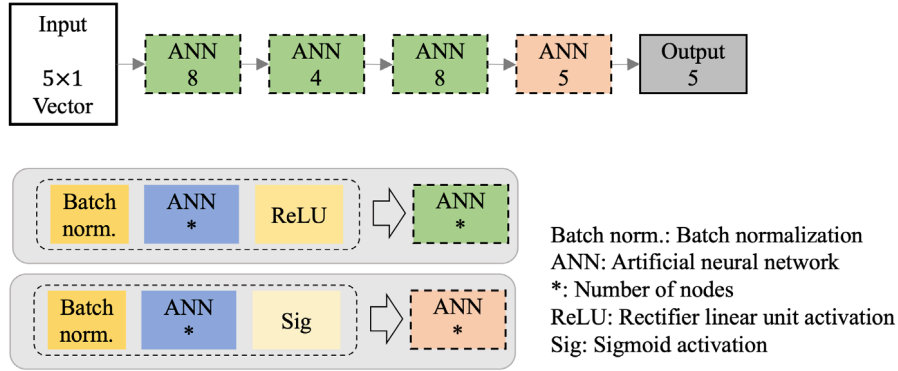

**Figure 6.** Detailed architecture of the DNN model

Figure 7 shows the history of convergence ratio across the iterations, which indicates that the fluctuations decrease as the number of data points used in training increases. Note that the introduction of stricter values may yield a more stable convergence trend. Using 66 runs of structural analysis (10 initial points + 56 points from the algorithm), it is possible to determine the noteworthy scenarios as listed in Table 3. Moreover, Table 4 presents the predicted possible noteworthy scenarios identified using the DNN model predictions. The results are consistent with the other methods proposed in the previous subsections. Since we use $R^*$ instead of $R$ for the upper limit of the search space, some scenarios that are located near the resilience threshold are also included in the list even though they satisfy the resilience criterion. Compared to the $n$-ball sampling method that requires numerous structural analyses, the surrogate model-based adaptive sampling can significantly mitigate the computational burden by focusing on exploring the domain near the failure points. Note that similarly to the preceding methods, the shaded cells in Tables 3 and 4 need not be considered during the redundancy analysis.

**Table 3.** Noteworthy scenarios identified by the adaptive sampling

| Case | Disruption scenarios ($F_i$) | $\beta_i$ | $\pi_i$ | $-\Phi^{-1}(\Phi(-\beta_i)\Phi(-\pi_i))$ |
|---|---|---|---|---|
| 1 | $\{C_1\overline{C_2}C_3\overline{C_4}C_5\}$ | 3.17 | −0.01 | 3.36 |
| 2 | $\{C_1\overline{C_2}C_3C_4C_5\}$ | 3.44 | 0.002 | 3.62 |
| 3 | $\{C_1\overline{C_2}C_3C_4\overline{C_5}\}$ | 1.68 | −0.002 | 1.99 |
| 4 | $\{\overline{C_1}C_2C_3\overline{C_4}C_5\}$ | 2.17 | 2.57 | 3.79 |
| 5 | $\{\overline{C_1}C_2C_3C_4\overline{C_5}\}$ | 2.52 | 2.91 | 4.26 |
| 6 | $\{\overline{C_1}C_2C_3\overline{C_4}C_5\}$ | 3.35 | 2.67 | 4.67 |
| 7 | $\{C_1\overline{C_2}C_3C_4\overline{C_5}\}$ | 4.11 | −0.005 | 4.26 |
| 8 | $\{C_1\overline{C_2}C_3C_4\overline{C_5}\}$ | 4.43 | −3.60 | 4.43 |

**Table 4.** Noteworthy scenarios predicted by the trained DNN model from the adaptive sampling

| Case | Disruption scenarios ($F_i$) | $\beta_i$ | $\pi_i$ | $-\Phi^{-1}(\Phi(-\beta_i)\Phi(-\pi_i))$ |
|---|---|---|---|---|
| 1 | $\{C_1\overline{C_2}C_3\overline{C_4}C_5\}$ | 3.17 | −0.01 | 3.36 |
| 2 | $\{C_1\overline{C_2}C_3C_4\overline{C_5}\}$ | 3.44 | 0.002 | 3.62 |
| 3 | $\{C_1\overline{C_2}C_3C_4\overline{C_5}\}$ | 1.68 | −0.002 | 1.99 |
| 4 | $\{\overline{C_1}C_2C_3\overline{C_4}C_5\}$ | 2.17 | 2.57 | 3.79 |
| 5 | $\{\overline{C_1}C_2C_3C_4\overline{C_5}\}$ | 2.52 | 2.91 | 4.26 |
| 6 | $\{\overline{C_1}C_2C_3\overline{C_4}C_5\}$ | 3.35 | 2.67 | 4.67 |
| 7 | $\{C_1\overline{C_2}C_3C_4\overline{C_5}\}$ | 4.11 | −0.005 | 4.26 |
| 8 | $\{\overline{C_1}C_2C_3C_4\overline{C_5}\}$ | 3.73 | −3.37 | 3.73 |

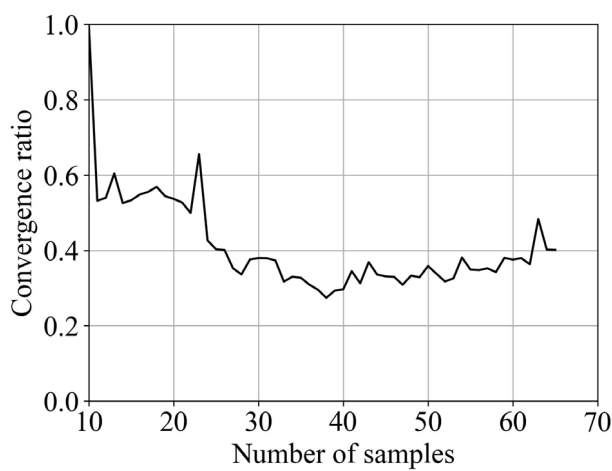

**Figure 7.** History of convergence ratio

3.5. *Comparison of the computational time*

To highlight the efficiency of the proposed methods, Table 5 summarizes the number of required simulations to identify the reliability indices of all critical scenarios using each method. The overall computational time is additionally presented where the analysis was performed on a personal computer equipped with Intel® i7-1360P. The results are compared with the brute force simulation method.

Note that there are two ways of estimating reliability indices in a brute force manner. First is to perform a large batch of MCS and estimate the reliability index of all scenarios at once through a postprocessing, i.e., using the occurrence rate of $F_i$ as its estimated probability. Another approach is to enumerate each initial disruption scenarios with advanced (e.g., adaptive) reliability algorithms. Depending on the resilience threshold level of interest and the number of disruption scenarios of the system, the preference may change.

In this example, the latter is selected because, as the highest reliability index among the initial disruptions scenarios is analytically found to be 8, it is infeasible to compute such rare probability through the MCS. Thus, a cross-entropy-based adaptive importance sampling using a Gaussian mixture (CE-AIS-GM) (Kurtz & Song, 2013) is employed. In other words, the importance sampling technique is employed to calculate the reliability indices of each initial scenarios for the brute force simulation method. The exact same computation is performed to estimate the reliability indices of the noteworthy scenarios identified by the proposed three methods. Consistently for all implementations, 100,000 sample points are used per adaptive iteration with 3 mixture models. The importance sampling density is considered converged when the c.o.v of the failure probability reaches 0.1.

**Table 5.** Computational costs for identifying noteworthy scenarios of the double-layer Daniels system and calculating their reliability indices

| Metrics | Brute force method | Sequential search method | $n$-ball sampling method | Surrogate model-based adaptive sampling |
|---|---|---|---|---|
| Number of simulations | $4.53 \cdot 10^7$ | $4.45 \cdot 10^6 + 2.5 \cdot 10^6$ | $10^4 + 3.0 \cdot 10^6$ | $66 + 4.3 \cdot 10^6$ |
| Computational time (minutes) | 77 | 11.9 | 5.1 | 7.3 |

In Table 5, for the proposed methods, the number of simulations is split into two terms, the first term standing for the required simulations to identify the noteworthy scenarios, and the second term representing the number of simulations required to estimate the reliability indices of the identified scenarios. Owing to the proposed methods, the computational burden is drastically reduced.

The sequential search method, due to its thorough search process, necessitates the largest number of simulations among the three methods. However, its foundation on rigorous mathematical derivation highlights its robustness, particularly for high-dimensional problems. This robustness will be explicitly demonstrated in Section 4.2.

**4. Numerical Investigations**

Four numerical examinations have been conducted in this section to demonstrate the applicability and effectiveness of the proposed pre-screening methods. First, the significance of identifying noteworthy scenarios is explored for varying resilience thresholds. Second, a truss bridge with a substantial number of random variables is introduced. Only the sequential search method is employed in this example due to the high dimensionality of the problem. The third examination aims to emphasize the advantage of the sampling scheme requiring fewer samples compared to the sequential search method. This is achieved by employing the *n*-ball sampling method to estimate noteworthy scenarios. However, the *n*-ball sampling method faces limitations when applied to problems demanding substantial computational resources per analysis. Thus, the fourth example employs the adaptive algorithm to identify noteworthy scenarios. Note that for the first, third and fourth examples, all three methods are employed to facilitate a comprehensive comparison of their computational costs.

The efficiency of the proposed methods is also compared with the brute force simulation approach. Among the two brute force approaches discussed in Section 3.5, the former one that relies on a batch MCS simulation is selected as it required less computational effort than the enumeration approach. Recall that the required number of samples $N$ in MCS to calculate the target probability $P_{target}$ with a given level of c.o.v $\eta$ is estimated as

$$N = \frac{1 - P_{target}}{\eta^2 \cdot P_{target}} \tag{8}$$

In order to ensure that all noteworthy scenarios are found and their reliability indices are estimated with sufficient precision, the target probability is set according to the resilience threshold level, i.e., $P_{target} = P_{dm}/(\lambda_H N_F)$. Additionally, the target c.o.v level of 0.05 is used for all numerical examples in this section. MCS is carried out for calculated $N$ through Eq. (8). The comparison of efficiency is made for the number of simulations and computational time taken to estimate the reliability index.

4.1. Computational costs along with different levels of resilience threshold

In the first example, the efficiency of the proposed methods is investigated for different resilience thresholds. Adopting an example in Lim et al., (2022), we examine a one-layer Daniels system with 6 bars sharing identical structural properties as shown in Figure 8. The yield stress of each bar $S_i$, *i*=1,2,…,6 is considered as random variables with a lognormal distribution with mean 400 MPa and c.o.v. of 0.35 under statistically independent condition. Note that the cross-sectional area is set as 1.0mm². A deterministic vertical load of 1,200kN is applied at the bottom and is equally distributed across the elements in the system. The component failure is defined in terms of the limit state function of $G(S_i) = S_i - 1,200/6 \leq 0$, *i*=1,2,….,6, while the system failure is defined as cascading failures of one or more elements.

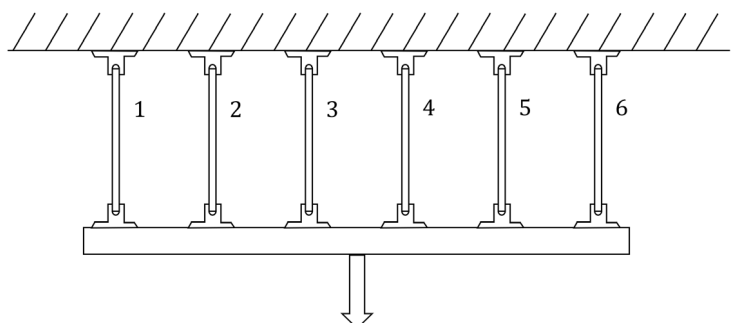

**Figure 8.** Single-layer Daniels system

The reference solution can be derived analytically in a closed from. However, to compare the efficiency of the proposed and the naïve Monte Carlo approach, the indices $\beta$ and $\pi$ are estimated using simulation-based approaches supposing that the analytic solution is not available. The analysis is repeated for six different resilience threshold levels: $10^{-2}$, $10^{-3}$, $10^{-4}$, $10^{-5}$, $10^{-6}$ and $10^{-7}$.

In Figure 9, the blue circles represent the reference analytic solution of the reliability and redundancy indices, while the black contours indicate the resilience thresholds. Due to polysymmetrical nature of the problem, multiple reference markers overlap. However, in the sampling-based predictions, we treat the model as a complete black-box and does not take any computational advantage from such symmetricity. As a result, the points estimated by the proposed method (or MCS) does not perfectly overlap exhibiting sampling variability.

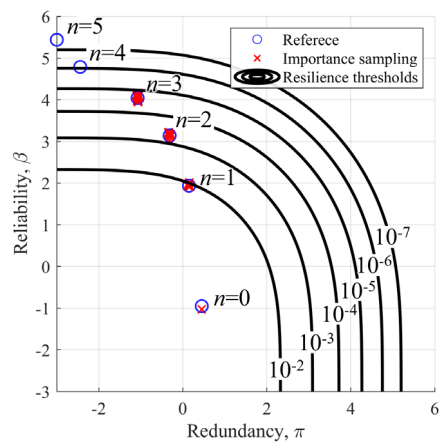

**Figure 9.** $\beta$–$\pi$ diagram of the single-layer Daniels system

The three proposed methods are employed to figure out the noteworthy scenarios, and the number of identified scenarios is summarized in Table 6. For the sequential approach, the joint component failure probabilities (e.g., $P(C_i)$) are identified using the cross-entropy-based adaptive importance sampling employing a von Mises-Fisher mixture (CE-AIS-vMFM) (Wang & Song, 2016). We used 500 samples per iteration stage and 3 mixture components to approximate the sampling density. In the $n$-ball sampling approach, the number of $n$-ball search samples is increased as the size of $n$-ball increases (i.e., as target resilience level gets stricter). We employ a heuristic increment factor that is proportional to the negative logarithmic scale for the increasement. Regarding the surrogate model-based algorithm, we used a DNN model architecture shown in Figure 6 and set the convergence ratio to 0.0001 to figure out failure domains near the high resilience thresholds, e.g., $10^{-7}$. The values of $\beta$ and $\pi$ for the identified noteworthy scenarios are estimated using the CE-AIS-vMFM under the same conditions. Note that other reliability analysis methods could be introduced in this regard. As shown in Figure 9 and Table 6, the number of noteworthy scenarios varies depending on the threshold level. Table 6 displays the number of "true" noteworthy scenarios and those identified by the proposed method. The numbers in the parentheses represent the number of actually critical scenarios among the identified noteworthy scenarios. Therefore, ideally this number should be identical to the ones shown in the reference column. Note that all method successfully identifies all the critical scenarios as noteworthy for the resilience target greater than 10[-6]. However, the surrogate model-based adaptive sampling approach struggles to properly identify several noteworthy scenarios for the extreme resilience threshold of $10^{-7}$. This deficiency arises from the limitations of the introduced surrogate model, which fails to accurately discern patterns among more than 50 failure domains in both input and output spaces. Addressing this issue will require further examination, potentially through the adoption of different surrogate models or the optimization of hyperparameters for the adaptive algorithm.

As previously mentioned, the proposed algorithms are intentionally conservative, often identifying a greater number of scenarios than the exact number. Among the methods, the sequential search algorithm has a smaller number of redundant scenarios, while the *n*-ball-based algorithms may additionally include many scenarios (that are located near the resilience threshold) as noteworthy. For example, for threshold of $10^{-4}$, the latter two methods designate almost twice more scenarios as noteworthy.

**Table 6.** The reference and estimated counts of noteworthy scenarios (inside parentheses is the number of reference noteworthy scenarios within the estimated counts)

| Resilience Threshold | Reference | Sequential search method | n-ball sampling method | Surrogate model-based adaptive sampling |
|---|---|---|---|---|
| $10^{-2}$ | 7 | 7 (7) | 7 (7) | 7 (7) |
| $10^{-3}$ | 7 | 9 (7) | 22 (7) | 20 (7) |
| $10^{-4}$ | 22 | 22 (22) | 42 (22) | 39 (22) |
| $10^{-5}$ | 42 | 42 (42) | 51 (42) | 47 (42) |
| $10^{-6}$ | 42 | 47 (42) | 58 (42) | 48 (42) |
| $10^{-7}$ | 57 | 57 (57) | 63 (57) | 54 (52) |

To inspect the impact of such conservatism on the overall efficiency, the overall number of model evaluations (including the simulations required to estimate *β*) is recorded and compared in Table 7. The number of MCS simulations are determined using Eq. (8). Similar to Table 5, the number is divided into two parts: the first term for the screening stage, and the second term for the estimation of the reliability indices of the identified scenarios. One can first notice that all the prescreening approach yields higher overall computational efficiency compared to brute force MCS, and as expected, the gain becomes higher as the threshold becomes stricter. As a price for finding more compact and complete set of noteworthy scenarios, the sequential search method requires more computational burden in the prescreening stage compared to the *n*-ball methods. Conversely, the *n*-ball methods get higher overall computational efficiency at the potential cost of chance of missing the noteworthy scenarios.

Even though *n*-ball method has shown satisfactory coverage throughout all the examples in the paper, due to its sampling variability and assumptions on the shape of limit state surface, it is possible to miss some critical scenarios. This trend is even more evident for surrogate model-based adaptive sampling, which shows superior overall efficiency, but was observed to miss critical scenarios when the probability scale is very high. Therefore, selection of the prescreening methods should be problem specific, and various a priori information and assumptions should be considered such as problem dimension, target probability scale, geometry of limit state surface, and model evaluation cost, as will be further illustrated in the following sections.

**Table 7.** Computational costs for identifying noteworthy scenarios of the single-layer Daniels system and calculating their reliability indices

| Threshold | Brute force MCS | Sequential search method | n-ball sampling method | Surrogate model-based adaptive sampling |
|---|---|---|---|---|
| $10^{-2}$ | $4 \cdot 10^4$ | $4.1 \cdot 10^4 + 1.0 \cdot 10^4$ | $2 \cdot 10^4 + 1.0 \cdot 10^4$ | $91 + 1.0 \cdot 10^4$ |
| $10^{-3}$ | $4 \cdot 10^5$ | $4.0 \cdot 10^4 + 1.4 \cdot 10^4$ | $3 \cdot 10^4 + 4.0 \cdot 10^4$ | $135 + 3.7 \cdot 10^4$ |
| $10^{-4}$ | $4 \cdot 10^6$ | $9.7 \cdot 10^4 + 4.0 \cdot 10^4$ | $4 \cdot 10^4 + 9.5 \cdot 10^4$ | $260 + 9.1 \cdot 10^4$ |
| $10^{-5}$ | $4 \cdot 10^7$ | $1.59 \cdot 10^5 + 9.7 \cdot 10^4$ | $5 \cdot 10^4 + 1.32 \cdot 10^5$ | $195 + 1.23 \cdot 10^5$ |
| $10^{-6}$ | $4 \cdot 10^8$ | $1.55 \cdot 10^5 + 9.7 \cdot 10^4$ | $6 \cdot 10^4 + 1.64 \cdot 10^5$ | $111 + 1.27 \cdot 10^5$ |
| $10^{-7}$ | $4 \cdot 10^9$ | $2.07 \cdot 10^5 + 1.56 \cdot 10^5$ | $7 \cdot 10^4 + 2.16 \cdot 10^5$ | $154 + 1.59 \cdot 10^5$ |

4.2. Truss bridge system analyzed using the sequential search method

A truss bridge subjected to two concentrated loads is considered to demonstrate the efficiency of the sequential search method. The configuration of the bridge system is presented in Figure 10 and structural properties, i.e., the Young's modulus and cross-sectional areas of the 25 members, are summarized in Table 8. A total of 27 random variables are introduced in this problem, including strength of the structural components and external loads, with their statistical information listed in Table 9. We assume a correlation coefficient of 0.6 between the external forces, while the other variables are considered to be independent of each other.

The failure of a component in this problem is defined as the compressive or tensile stress caused by an external load exceeding the yield strength. Once a structural member fails, it is assumed that its stiffness is reduced to 20% of its initial value, following (Li, 2006). In contrast, the system failure event is defined as the occurrence of structural instability in the truss bridge. When calculating the redundancy indices, load redistribution due to the initial distribution $F_i$ is taken into account, and the joint probability distribution of the random variables is updated to condition on $F_i$. More information on the distribution update is available in Yi & Kim (2023). The resilience threshold for this problem is set to $10^{-4}$.

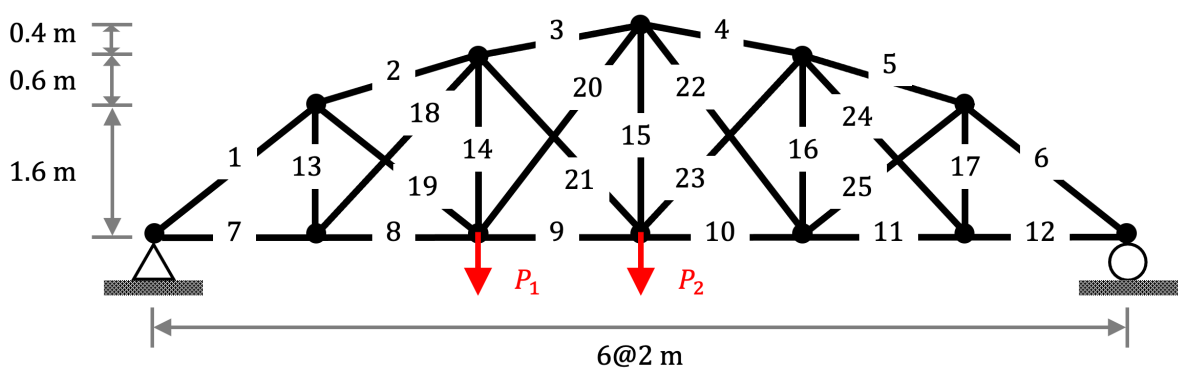

**Figure 10.** Configuration of truss bridge having 25 members

**Table 8.** Structural properties of the truss bridge

| Truss members | Cross-sectional area (m²) | Young's modulus (GPa) |
|---|---|---|
| 1, 6 | $15 \times 10^{-4}$ | |
| 2, 5 | $16 \times 10^{-4}$ | |
| 3, 4 | $18 \times 10^{-4}$ | 210 |
| 7 – 12, 15, 20 – 23 | $14 \times 10^{-4}$ | |
| 13, 17 – 19, 24, 25 | $13 \times 10^{-4}$ | |
| 14, 16 | $12 \times 10^{-4}$ | |

**Table 9.** Distribution types and statistical parameters of the random variables

| Random variables | Distribution type | Mean | Coefficient of variation |
|---|---|---|---|
| External forces ($P_1$ and $P_2$) | Lognormal | 160 kN | 0.1 |
| Yield strength | Normal | 276 MPa | 0.05 |

The numerical analysis revealed that the CE-AIS-GM method inadequately identifies the reliability index. In contrast, CE-AIS-vMFM (Wang & Song, 2016), designed to handle high-dimensional random variable spaces, showed good performance, in this regard. Moreover, it is remarked that both the *n*-ball sampling method and the surrogate model-based adaptive sampling algorithm do not provide promising results for this specific problem. This occurrence is attributed to the challenge of adequately covering the high-dimensional sample space (i.e., 27). Thus, only the results from the sequential search method are presented for this example.

Following the procedure outlined in Section 3.2, the reliability indices of a single component failure scenario are first estimated and summarized in Phase 1 of Figure 11. Note that to simplify the figure, component failure events with a reliability index greater than 10 are not shown. Truss members #3 and #9 are found to be most vulnerable to the external forces, which agrees with the intuition given the configuration of the structural system and the loads presented in Figure 10. By comparing with the given resilience threshold of $-\Phi^{-1}(10^{-4}) = 3.719$, it is found that the reliability index of 5 component failure events (truss members #1, #2, #3, #8, and #9) is smaller than the resilience target. Those events are marked with a solid circle in Figure 11. The intersection of two-event combinations which do not satisfy the resilience target is inspected in the next phase.

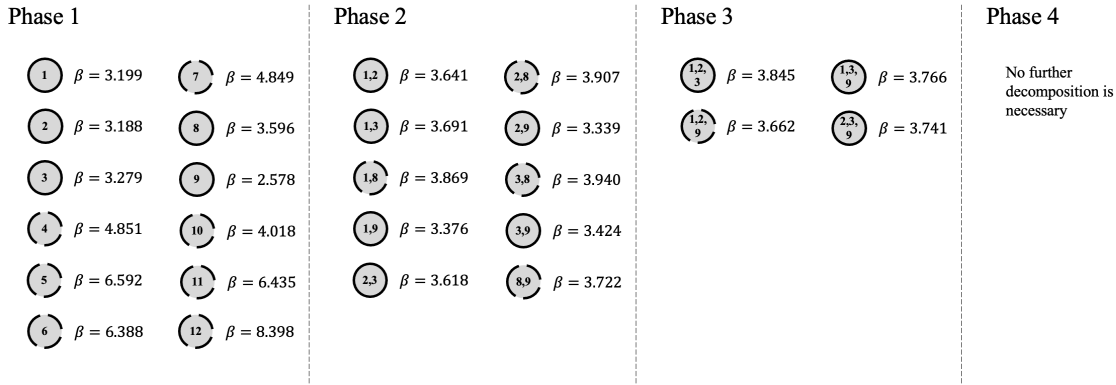

**Figure 11.** Identification of the noteworthy scenarios of the truss bridge example using the sequential search method

The identification of the components failure events whose reliability index is smaller than the resilience target is continued up to Phase 4, where no further decomposition is possible (see Figure 11). By collecting all the initial disruption scenarios whose components are not excluded during the process, 12 possible noteworthy initial disruption scenarios are identified and listed in Table 10. Such a number is $3.57 \cdot 10^{-5}\%$ of the total number of initial disruption scenarios. Note that, unlike Tables 1-4, due to the large number of structural components, only the failed components are denoted in the second column of Table 10 to represent the initial disruptions scenarios.

**Table 10.** Noteworthy failure scenarios of the truss bridge and estimated $\beta$ and $\pi$

| Case | Disruption scenarios ($F_i$) | $\beta_i$ | $\pi_i$ | $-\Phi^{-1}(\Phi(-\beta_i)\Phi(-\pi_i))$ |
|---|---|---|---|---|
| 1 | $\{C_1\}$ | 3.40 | 0.38 | 3.67 |
| 2 | $\{C_2\}$ | 3.36 | $-2.64$ | 3.36 |
| 3 | $\{C_3\}$ | 3.47 | $-3$ | 3.47 |
| 4 | $\{C_8\}$ | 3.89 | $-2.35$ | 3.89 |
| 5 | $\{C_9\}$ | 2.64 | $-0.95$ | 2.70 |
| 6 | $\{C_1 C_2\}$ | 4.07 | $< -3$ | 4.07 |
| 7 | $\{C_1 C_3\}$ | 4.17 | $< -3$ | 4.17 |
| 8 | $\{C_1 C_9\}$ | 3.56 | $-2.45$ | 3.56 |
| 9 | $\{C_2 C_3\}$ | 4.07 | $< -3$ | 4.07 |
| 10 | $\{C_2 C_9\}$ | 3.52 | $< -3$ | 3.51 |
| 11 | $\{C_3 C_9\}$ | 3.62 | 0.68 | 3.96 |
| 12 | $\{C_1 C_2 C_9\}$ | 3.89 | $< -3$ | 3.89 |

Then the reliability and redundancy indices of the noteworthy scenarios are estimated using a cross-entropy-based adaptive importance sampling using the CE-AIS-vMFM (Wang & Song, 2016). The number of samples used in each importance sampling is 1,000 with 3 mixture models. Eventually, 6 initial disruption scenarios are identified to have the resilience level lower than the target: Cases 1, 2, 3, 5, 8, and 10. Note that similarly to the Daniels system example in Section 3, it is unnecessary to calculate the values in the gray shaded area of Table 10. The estimated initial disruption scenarios are well aligned with those obtained by performing brute force MCS, as shown in the β–π diagram in

Figure 12. The "Case" denoted next to the corresponding β–π point in Figure 12 corresponds to the first column of Table 10.

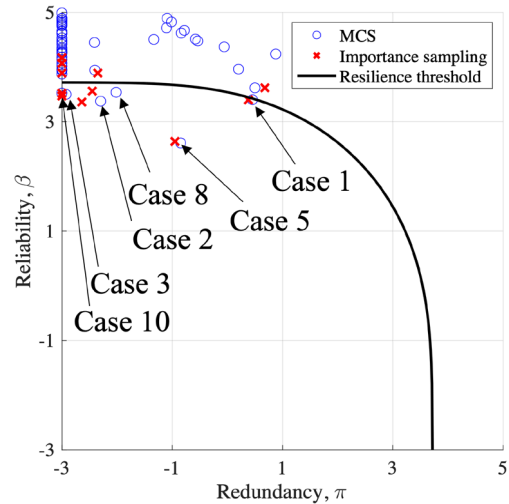

**Figure 12.** β–π diagram of the truss bridge system

Table 11 summarizes the comparison between the brute force MCS and the sequential search method with the CE-AIS-vMFM in terms of required simulation numbers and overall computational time. The first term of the number of simulations in the sequential search method denotes the number of structural analyses needed to identify the noteworthy scenarios, while the subsequent term represents the simulations required to compute the reliability indices. These numerical comparisons demonstrate the effectiveness of identifying noteworthy scenarios in reducing overall computational costs. Lastly, it is remarked that while the proposed method has shown satisfactory performance in this example with moderately high input dimension, its applicability to problems with an extremely large number of random variables, such as 100 or 200, needs further demonstration.

**Table 11.** Computational costs for identifying noteworthy scenarios of the truss bridge system and calculating their reliability indices

| Metrics | MCS | Sequential search method |
|---|---|---|
| Number of simulations | $4 \cdot 10^6$ | $1.2 \cdot 10^5 + 6.9 \cdot 10^4$ |
| Computational time (seconds) | 10.13 | 0.479 |

4.3. Truss building structure analyzed using the *n*-ball method

The *n*-ball sampling approach proposed in Section 3.3 is demonstrated for a truss building model subjected to a static load. The building model has 6 members as shown in Figure 13 and it is assumed that all the members have the same yield strength. The yield strength value and the applied loads are considered as random variables, as shown in Table 12. The 6 random variables are mutually independent. Other structural parameters are considered deterministic as presented in Table 13. Local damage is induced when the member stress exceeds the yield strength, and the stiffness is assumed to be reduced to half after the damage. System-level failure is induced when more than two members are damaged. The resilience threshold for this problem is set to $10^{-5}$. Following the procedure in Section 3.3, the radius of *n*-ball is set as $1.05R$, where $R = -\Phi^{-1}(10^{-5}) = 4.265$, and $10^4$ samples are generated uniformly across the *n*-ball volume.

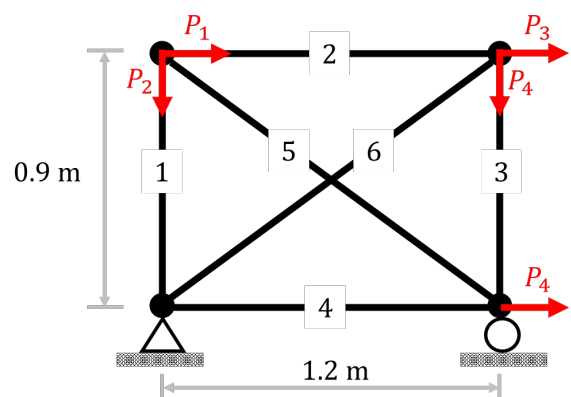

**Figure 13.** Truss building model subjected to horizontal and vertical loads

**Table 12.** Distribution types and statistical parameters of the random variables

| Random variables | Distribution type | Mean | Coefficient of variation |
|---|---|---|---|
| Horizontal forces ($P_1, P_3, P_5$) | Normal | 5 kN | 0.5 |
| Vertical forces ($P_2$ and $P_4$) | Normal | 15 kN | 0.5 |
| Yield strength | Normal | 276 MPa | 0.2 |

**Table 13.** Structural properties of the truss building structure

| Truss members | Cross-sectional area (m²) | Young's modulus (GPa) |
|---|---|---|
| 1 | $20 \times 10^3$ | |
| 2 | $5 \times 10^3$ | |
| 3 | $20 \times 10^3$ | 70.6 |
| 4 | $35 \times 10^3$ | |
| 5 | $20 \times 10^3$ | |
| 6 | $10 \times 10^3$ | |

As results, 8 noteworthy scenarios are identified, which are listed in Table 14. By comparing to the brute force MCS results obtained using $4 \times 10^7$ simulations as shown in Figure 14, one can see that all the noteworthy scenarios are well captured using a smaller number of simulations. Because the approach relies on random sampling, each trial may give slightly different noteworthy cases. To inspect the robustness of the method under the sampling variability, the analysis is repeated 100 times with different random seeds, and the proposed approach detected 7~9 noteworthy scenarios each time, and those always included the 6 critical cases that do not satisfy the resilience target. Note that, with the same number of samples ($10^4$), MCS was able to detect all 6 cases only in 6 out of 100 trials.

Subsequently, 8 independent subset simulations are performed (Li et al., 2019; Papaioannou et al., 2015) to obtain the reliability indices of the noteworthy scenarios, and the results are shown in the third column of Table 14. The values are obtained using a total of $3 \times 10^4$ simulations including the $10^4$ simulations used to identify the noteworthy scenario. Note that to obtain MCS results with similar accuracy, e.g., the target probability down to $P_f = 10^{-5}$ with a coefficient of variation of $\eta = 0.05$, one needs to perform simulation more than $N = 4 \times 10^7$ times according to Eq. (8), showing that the efficiency gain was more than a factor of 1000. This factor is expected to increase when the target probability (the resilience threshold) is lower, or the input dimension is smaller. However, the $n$-ball screening approach suffers from the curse of dimensionality when the input dimension gets higher. In such cases, it is recommended to use the sequential search approach.

The overall computational costs are compared with other methods and MCS, as presented in Table 15. All the three methods successfully identified the noteworthy scenarios and their reliability indices are estimated with significantly lower number of model evaluations compared to the brute force MCS.

**Table 14.** Noteworthy failure scenarios of the truss building structure and estimated β and π

| Case | Disruption scenarios ($F_i$) | $\beta_i$ | $\pi_i$ | $-\Phi^{-1}(\Phi(-\beta_i)\Phi(-\pi_i))$ |
|---|---|---|---|---|
| 1 | $\{C_1C_2C_3C_4C_5C_6\}$ | −2.51 | 3.50 | 3.50 |
| 2 | $\{C_1\overline{C_2}C_3C_4C_5C_6\}$ | 3.52 | 0.88 | 3.94 |
| 3 | $\{\overline{C_1}C_2C_3C_4C_5C_6\}$ | 2.77 | 1.04 | 3.34 |
| 4 | $\{\overline{C_1}C_2C_3C_4\overline{C_5}C_6\}$ | 4.11 | −0.05 | 4.27 |
| 5 | $\{C_1\overline{C_2}C_3\overline{C_4}C_5C_6\}$ | 3.60 | $< -3$ | 3.60 |
| 6 | $\{C_1\overline{C_2}C_3C_4\overline{C_5}C_6\}$ | 4.08 | $< -3$ | 4.07 |
| 7 | $\{\overline{C_1}C_2C_3\overline{C_4}C_5C_6\}$ | 3.72 | $< -3$ | 3.73 |
| 8 | $\{C_1\overline{C_2}\overline{C_3}C_4C_5C_6\}$ | 3.92 | $< -3$ | 3.92 |

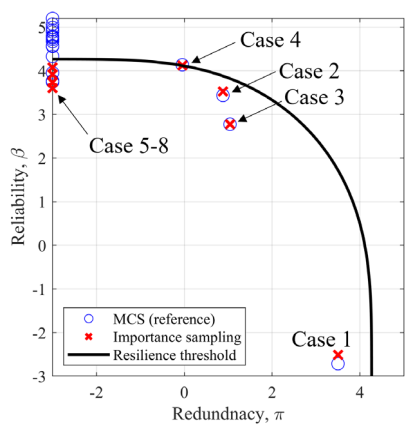

**Figure 14.** β–π diagram of the truss building structure

**Table 15.** Computational costs for identifying noteworthy scenarios of the truss building structure and calculating their reliability indices

| Metrics | MCS | Sequential search method | $n$-ball sampling method | Surrogate model-based adaptive sampling |
|---|---|---|---|---|
| Number of simulations | $4 \cdot 10^7$ | $6.3 \times 10^4 + 1.4 \times 10^4$ | $1.0 \times 10^4 + 2.0 \times 10^4$ | $130 + 5.0 \times 10^4$ |
| Computational time (minutes) | 822.0 | 43.5 | 11.1 | 26.7 |

4.4. 3-story building structure analyzed using the surrogate model-based adaptive algorithm

The third example investigates a three-story, four-bay building model subjected to lateral static forces shown in Figure 15. The statistical parameters of the external forces (red arrows) are summarized in Table 16. Note that the shape of the lateral forces mimics the second mode shape of the structure that induces diverse initial disruption scenarios. The numerical model used in the study is based on the SAC joint venture project and is developed using OpenSees software (McKenna, 2011). The dimensions of each bay and story are 9.15 m and 3.96 m, respectively. A fiber section is employed for both beams and columns made of "Steel 01" material, with a material stiffness of 200 GPa, and yield strength of 248 MPa and 345 MPa for the beam and column, respectively. A post-yield stiffness-to-initial stiffness ratio of 1% is assumed. Each story includes a weak column on the rightmost side, and a rigid diaphragm assumption is made.

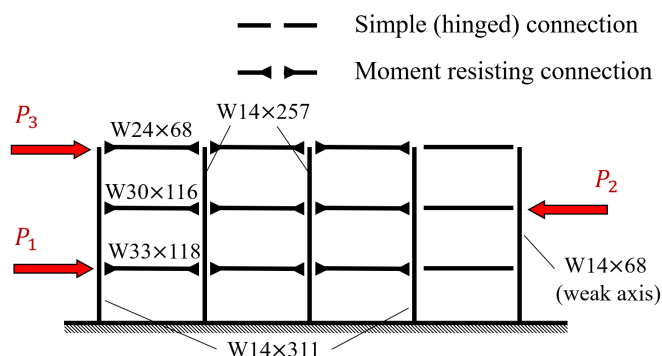

**Figure 15.** 3-story steel building model subjected to lateral forces

**Table 16.** Statistical parameters of the external loads

| External forces | Mean (kN) | Coefficient of variation |
|---|---|---|
| $P_1$ | 3,000 | 0.25 |
| $P_2$ | 4,000 | 0.20 |
| $P_3$ | 5,000 | 0.15 |

We adopt the initial disruption scenarios described in Yi & Kim (2023), which consists of the failure of the weakest columns (i.e., the rightmost columns) in each story. This results in a total of $2^3$ initial disruption events. The limit-state for component failure occurs when the stress exceeds the yield stress, while the system-level limit-state is defined as the roof drift exceeding the maximum allowable roof drift, which is assumed to be 10% in this example. If a structural component fails, the stiffness of the weak column is assumed to be reduced to 50% to account for the performance degradation of the component and the corresponding load redistribution. It should be noted that the proposed methods are not limited to the illustrated limit-state functions and expressions of damaged components, and alternative modeling details can be employed.

By setting the resilience threshold of $10^{-4}$, the resilience analysis is conducted to identify noteworthy scenarios using the same hyperparameters of the adaptive algorithm and the DNN architecture employed in Section 3.4. Note that the input of the DNN model is the external forces transformed to the uncorrelated standard normal space **u**, while the output is a 3 × 1 binary vector representing the failure (or safe) status of the three components. Figure 16 presents the history of convergence ratio across the iterations. By performing 53 structural analysis (6 initial points + 47 points from the algorithm), 3 noteworthy scenarios are identified as shown in Table 17, where $C_i$ represents the *i*-th story weak column failure event. Moreover, it is found that the possible noteworthy scenarios obtained from the DNN model are equivalent to those in Table 17. The corresponding reliability and redundancy indices are estimated using a CE-AIS-GM developed by Kurtz & Song (2013). The number of samples used in the importance sampling is 1,000 with 2 mixture models.

Table 17 shows that Cases 1 and 2 are the failure scenarios that do not meet the resilience target of $-\Phi^{-1}(10^{-4}) = 3.72$, which is well aligned with those obtained from brute force MCS as presented in Figure 17. Moreover, to demonstrate the robustness of the surrogate model-based adaptive algorithm, the analysis is carried out 100 times with different random seeds. The proposed methods consistently identify the three cases in Table 17 as noteworthy scenarios. Nonetheless, due to the method's dependency on sampling within the random variable space, the number of required simulations varies for each random seed. Figure 18 illustrates a box plot depicting the distribution of the required number of simulations to identify the noteworthy scenarios for different random seeds. The median value of the required number of simulations is estimated as 54.

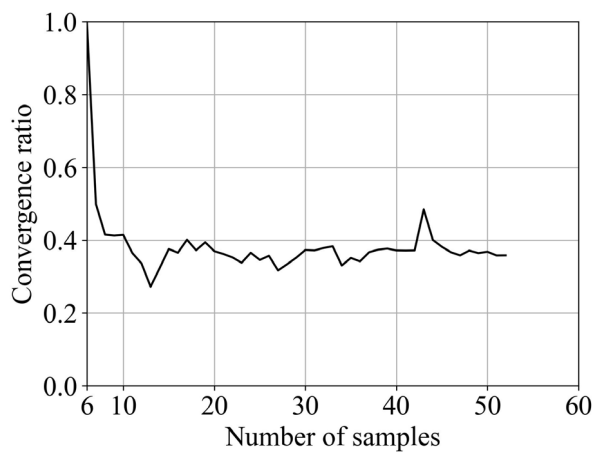

**Figure 16.** Convergence ratio as the iteration proceeds

**Table 17.** Noteworthy failure scenarios of the 3-story building model identified by adaptive sampling

| Case | Disruption scenarios ($F_i$) | $\beta_i$ | $\pi_i$ | $-\Phi^{-1}(\Phi(-\beta_i)\Phi(-\pi_i))$ |
|---|---|---|---|---|
| 1 | $\{C_1\overline{C_2}C_3\}$ | 1.20 | $-8.11 \cdot 10^{-4}$ | 1.57 |
| 2 | $\{C_1\overline{C_2}C_3\}$ | 3.00 | $-1.33$ | 3.03 |
| 3 | $\{\overline{C_1C_2}C_3\}$ | 3.38 | 2.02 | 4.32 |

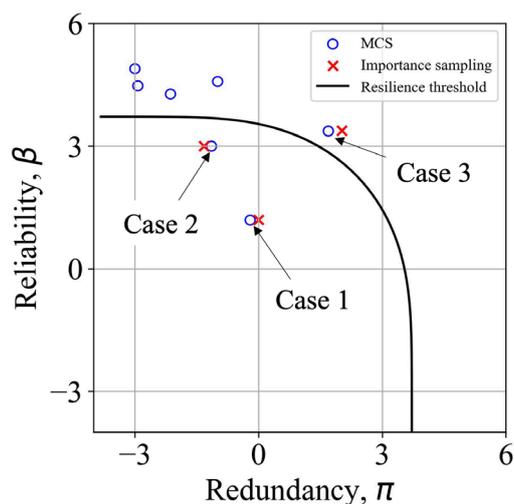

**Figure 17.** β–π diagram of the 3-story building model obtained by brute force MCS

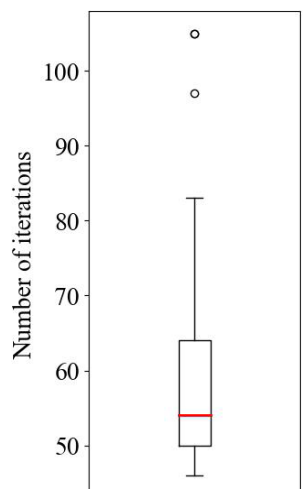

**Figure 18.** Box plot of the number of iterations for the adaptive algorithm

To further highlight the effectiveness of the surrogate model-based adaptive method, the computational costs and number of simulations are compared with other methods and MCS, as presented in Table 18. All the three methods successfully identified the noteworthy scenarios. Similar to the previous one, the first term of the number of simulations in the proposed three methods denotes the number of structural analyses needed to identify the noteworthy scenarios, while the subsequent term represents the simulations required to compute the reliability index.

Given the high computational cost associated with structural analysis using OpenSees, the superiority of the surrogate model-based adaptive sampling becomes evident. These compelling results demonstrate that the proposed algorithm combined with the advanced reliability analysis techniques provides a robust and reliable framework for identifying noteworthy scenarios by efficiently exploring the rare event domains. However, since the component and structural limit-state functions in the numerical investigations incorporate a single performance equation, the applicability to problems having multiple limit-states needs to be further elaborated.

**Table 18.** Computational costs for identifying noteworthy scenarios of the 3-story building model and calculating their reliability indices

| Metrics | MCS | Sequential search method | $n$-ball sampling method | Surrogate model-based adaptive sampling |
|---|---|---|---|---|
| Number of simulations | $4 \cdot 10^6$ | $1.6 \cdot 10^4 + 1.1 \cdot 10^4$ | $10^4 + 1.1 \cdot 10^4$ | $53 + 1.1 \cdot 10^4$ |
| Computational time (minutes) | 4,224 | 28.5 | 22.2 | 11.7 |

## 5. Conclusions

The aim of this study was to propose methods that can reduce the computational burden for estimating the reliability and redundancy indices within the context of system-reliability-based disaster resilience analysis. Three methods were developed to accelerate the resilience analysis by identifying noteworthy initial disruption scenarios: the sequential search method, the $n$-ball sampling method, and the surrogate model-based adaptive sampling algorithm. The efficiency of the proposed approaches was demonstrated through numerical examples including a bridge structure with a large number of random variables and real-world building models with substantial computational demands. The required number of simulations and computational costs were compared with the brute force Monte Carlo simulation (MCS) to demonstrate the computational benefit of identifying the noteworthy scenarios. Moreover, the results were comprehensively compared within the three proposed methods, which emphasized the merit of each approach. For instance, in the truss bridge structure having 25 random variables, the sequential search method reduced the number of scenarios that should be covered for the resilience analysis from $2^{25}$ to 12, which is $3.57 \cdot 10^{-5}$%. It is remarked that, although the assumptions we used to model the damaged component and system level behavior and was based on literature, introduction of a more realistic damage model is desired to further facilitate the applicability of system-reliability-based disaster resilience analysis in practice.

One possible extension of this work could be to apply the proposed methods to structural systems subjected to stochastic excitations. By combining the methods proposed in this study with our previous work (Yi & Kim, 2023), it is possible to develop novel approaches that assess the resilience performance of structures while considering the variability in random excitations. Additionally, as highlighted in the Introduction, resilience analysis involves a comprehensive assessment of structural systems, encompassing decision-making and time-dependent restoration. Further research is needed to interconnect the proposed methods with the restoration process, thus establishing the proposed methods as a key element in the "resilience" analysis framework.

**Acknowledgment:** This research was supported by the National Research Foundation of Korea (NRF) grant funded by the Korea government (MSIT) (RS-2023-00242859).